\documentclass[aps,epsfig,showpacs]{revtex4}
\usepackage{graphicx}
\usepackage{amsmath}
\usepackage{latexsym}
\usepackage{amsfonts}
\usepackage{amssymb}
\usepackage{dsfont}
\usepackage{array}
\usepackage{color}
\usepackage{subfigure}
\usepackage{verbatim}
\newcommand{\ket}[1]{\left\vert#1\right\rangle}
\newcommand{\bra}[1]{\left\langle#1\right\vert}

\begin{document}
\newcommand{\Q}[1]{{\color{red}#1}}
\newcommand{\blue}[1]{{\color{blue}#1}}
\newcommand{\red}[1]{{\color{red}#1}}
\newcommand{\Change}[1]{{\color{green}#1}}

\title{Deterministic nonlinear gates with oscillators mediated by a qubit}

\author{
Kimin Park}
\email{park@optics.upol.cz}
\affiliation{
Department of Optics, Palack\'y University, 17. listopadu 1192/12, 77146 Olomouc, Czech Republic}
\author{Petr Marek}
\author{Radim Filip}
\affiliation{
Department of Optics, Palack\'y University, 17. listopadu 1192/12, 77146 Olomouc, Czech Republic}
\date{\today}

\begin{abstract}
Quantum nonlinear operations for harmonic oscillator systems play a key role in the development of analog quantum simulators and computers. Since a variety of strong highly nonlinear operations are unavailable in the existing physical systems, it is a common practice to approximate them by using conditional measurement-induced methods. The conditional approach has several drawbacks, the most severe of which is the exponentially decreasing success rate of the strong and complex nonlinear operations. We show that by using a suitable two level system sequentially interacting with the oscillator, it is possible to resolve these issues and implement a nonlinear operation both nearly deterministically and nearly perfectly. We explicitly demonstrate the approach by constructing self-Kerr and cross-Kerr couplings in a realistic situation, which require a feasible dispersive coupling between the two-level system and the oscillator.
\end{abstract}
\flushbottom
\maketitle
\thispagestyle{empty}

\section*{Introduction}
Quantum computers or quantum Turing machines~\cite{quantumcomputer} take advantage of their quantum mechanical architecture and are capable of solving tasks which are exponentially hard for their classical counterparts \cite{algorithm}.
Their predecessors are quantum simulators \cite{simulators,Feynman1982}, which seek to emulate specific quantum dynamics of particular quantum systems in place of general processing. The fundamental principle of the simulations relies on mapping the complex quantum systems onto other more accessible and better controllable ones, such as trapped ions \cite{simulation_ions}, photons \cite{simulation_photons}, atomic lattices \cite{atomiclattice} and superconducting circuit \cite{superconducting}.
The analog simulators are dedicated to continuous variables (CV) systems with infinite dimensional Hilbert space~\cite{CVsystems}. These systems allow for simulations of unexplored highly nonlinear open quantum dynamics \cite{FilipPRA2005MeasCVinter, MiwaPRL2014,LloydPRL1999, GKP, MarekPRA2011Cubic, SefiPRA2013, MiyataPRA2016}.
  Some CV nonlinear operations naturally appear in other physical systems, such as Bose-Einstein condensates \cite{BEC}, cold ions \cite{cold_ions}, or circuit quantum electrodynamics \cite{KirchmairNat2013Kerr}. The spectrum of nonlinear operations is however limited and typically determined by the unique physics of specific experimental platforms.

A broader set of nonlinear operations for quantum harmonic oscillator can be elegantly realized by coupling them to suitable two-level systems (qubits) \cite{LeibfriedRMP2003Trappedions, XiangRMP2013hybridsuperconducting, AspelmeyerRMP2014CavityOptomechanics, ReisererRMP2015CavityNetwork, LodahlRMP2015PhotonicNanostructures}. This realization is possible because the two-level systems are naturally nonlinear due to their saturability and offer a wide variety of qubit-oscillator couplings. The nonlinear nature in turn leads to dynamics of the oscillator which can be used for deterministic generation of nonclassical states \cite{NCstates} or for conditional realization of nonlinear quantum potentials \cite{Park2016Rabi,ParkPRA2016JC}. The two level systems are also beneficial from a technical standpoint, allowing for a significantly larger number of individual interactions \cite{SayrinNat2011atom} than what is allowed for purely optical ancillary single photon states \cite{FiurasekPRA2009, ParkPRA2014Xgate}. The conditional nature of these hybrid operations, however, limits them in their suitability for practical applications as well as quantum simulations, which ultimately leads to success rate exponentially decreasing with the number of operations involved.

In this report we propose a method for deterministic implementation of nonlinear unitary operations for quantum harmonic oscillators sequentially coupled to single qubits. This method relies on employing a sequence of available non-commuting qubit-oscillator interactions, similarly as in \cite{LloydPRL1999,LloydARXIV2000Hybrid,SefiPRL2011, SpillerNJP2006}. The qubits act only as mediators rather than for control unlike the conceptually similar quantum Zeno gates \cite{HuangPRA2008}, starting and finishing the operation in a factorized state.
The repeated gates incrementally create a Zeno-like nonlinear unitary dynamics deterministically and with a nearly unit fidelity.
We illustrate the quality of the proposed method by explicitly analyzing realization of the self-Kerr and cross-Kerr nonlinearities done with help of a qubit sequentially coupled to the oscillator by dispersive interactions \cite{BlaisPRA2004cQEDsupcond,GleyzesNat2007dispersiveatom,GuerlinNat2007Dispersiveatom,SchusterNat2007Superconducting,ThompsonNat2008mechanical,JohnsonNatPhys2010QND} under photon losses.

\section*{Short-time oscillator interaction transduced by a qubit}

\begin{figure}[ht]
\centering
\includegraphics[width=350px]{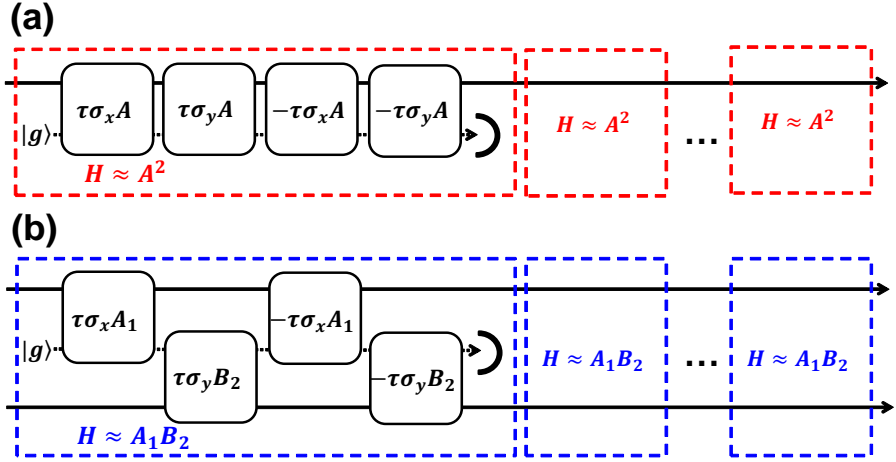}
\caption{Concept of deterministic gates with oscillators mediated by a qubit where the interactions $H\approx\sigma_{x,y}A$, $H\approx\sigma_{x,y}A_1$ and $H\approx\sigma_{x,y}B_2$ between optical mode and ancillary mode are arranged to achieve a high-order nonlinearity: (a) scheme for single-mode optical interaction operator, and (b) scheme for a two-mode optical evolution operator.  Each box with a written interaction Hamiltonian $H$ represents the evolution $\exp[i H]$ for a unit of time, and the different colored boxes represent different operators acting on the ancillas. The ancillas prepared in a chosen state $\ket{g}$ are discarded after each set of interactions. Repetition of these unit approximate operators represented by dashed boxes makes high-strength operators.  }
\label{Setup}
\end{figure}
Let us start by considering a short time evolution of a quantum oscillator mediated by a single qubit. The unitary oscillator-qubit interaction that enables the desired dynamics is governed by Hamiltonians of the type $H_{\hat{A}}=\hbar\hat{\sigma}_j\hat{A}$, where $\hat{\sigma}_j$ with $j = x,y,z$ relates to the qubit system and stands for one of Pauli matrices, and $\hat{A}$ is an operator acting on the oscillator. We assume resonant cases when free evolution Hamiltonians can be eliminated. To achieve the desired gate on the oscillator, we can consider a pair of non-commuting unitary operators $\hat{U}_x=\exp[i\tau\hat{\sigma}_x \hat{A}]$ and $\hat{U}_y=\exp[i\tau\hat{\sigma}_y \hat{B}]$ where the oscillator operators $\hat{A}$ and $\hat{B}$ commute $[\hat{A},\hat{B}] = 0$. As depicted in Fig.~\ref{Setup}a, we can join them into a sequence $\hat{U}_{xyxy}=\hat{U}_{x}\hat{U}_y\hat{U}_x^\dagger\hat{U}_y^\dagger$ following the idea of geometric phase effect~\cite{berry}. In a manner similar to \cite{LloydPRL1999,Loock2010}, this operator can be simplified to
\begin{align}
&\hat{U}_{xyxy}=\exp[i\tau\hat{\sigma}_x \hat{A}]\exp[i\tau\hat{\sigma}_y \hat{B}]\exp[-i\tau\hat{\sigma}_x \hat{A}]\exp[-i\tau\hat{\sigma}_y \hat{B}]\nonumber\\
&=1-2\sin^2[\tau\hat{A}]\sin^2[\tau\hat{B}]+i\sin[2\tau \hat{A}]\sin^2[\tau\hat{B}]\hat{\sigma}_x-i\sin^2[\tau \hat{A}]\sin[2\tau\hat{B}]\hat{\sigma}_y-\frac{i}{2}\sin[2\tau \hat{A}]\sin[2\tau\hat{B}]\hat{\sigma}_z\approx\exp[-2i\tau^2\hat{\sigma}_z \hat{A} \hat{B}]\equiv \hat{U}_{\hat{\sigma}_z \hat{A}\hat{B}},
\label{eq:mix}
\end{align}
where the last line corresponds to a weak strength limit $\tau\ll 1$~\cite{supp}. The resulting oscillator dynamics is driven by the product of operators $\hat{A}\hat{B}$ and coupled to the qubit by $\hat{\sigma}_z$. The qubit degree of freedom can be  straightforwardly eliminated by preparing and measuring the qubit system in one of the relevant eigenstates, such as $|g\rangle$. The measurement then substitutes the discarding of qubit depicted in Fig.~\ref{Setup}a. The whole sequence $\bra{g}\hat{U}_{xyxy}\ket{g}$ then realizes a {\em conditional} operator
\begin{align}
   &\hat{O}_1=\bra{g}\hat{U}_{xyxy}\ket{g}=1-2\sin^2[\tau\hat{A}]\sin^2[\tau\hat{B}]+i\sin[2\tau \hat{A}]\sin[2\tau \hat{B}]/2,\label{operation_O1}
\end{align}
which approximates unitary operation
 \begin{align}
 U_{\hat{A}\hat{B}}=\exp[-2i\tau^2\hat{A}\hat{B}]\label{eq:target}
 \end{align}
in the limit of small $\tau$. The commutativity of $\hat{A}$ and $\hat{B}$ restricts the generality of the scheme, but still allows for many interesting cases. The base operators $\hat{A}$ and $\hat{B}$ can be compatible operators on a single oscillator (as in Fig.\ref{Setup}a), or different operations on two separate oscillators (illustrated in Fig.\ref{Setup}b). The most apparent scenarios in which the product of two operators is highly nontrivial and practically useful operation are the self-Kerr and cross-Kerr evolutions, which we will address in detail later.

\section*{Near-unitarity of short-time realistic interaction}

The perfect operation (\ref{eq:target}) is realized only in the limit of short time $\tau \rightarrow 0$. However, we can increase the strength by repeating the individual operations.
In each step, the ancillary qubit is initialized in the ground state, led to interact with the oscillator systems, and finally projected onto the ground state again. It does not matter whether  a single physical qubit is used repetitively or if a number of different systems is employed. In any case, $R$ repetitions realize quantum operation $\hat{O}_R = (\hat{O}_1)^R$ which approximates the ideal operation $ \hat{O}_T\equiv e^{-2iR\tau^2 \hat{A} \hat{B}}$. Interestingly enough, in the limit of sufficiently small $\tau$ the re-initialization of qubit is not needed, as the approximate operator can be also obtained as $\hat{O}_R = \bra{g}(\hat{U}_{xyxy})^R\ket{g}$.

For a specific test state $\ket{\psi}$, the performance of the operation can be quantified by looking at its successful implementation probability $P_s=\bra{\psi}\hat{O}_R^\dagger \hat{O}_R\ket{\psi}$  and fidelity $F=|\bra{\psi}\hat{O}_T^\dagger \hat{O}_R\ket{\psi}|^2/P_s$. These metrics inherently depend on the chosen state $\ket{\psi}$, but we can also directly analyze the sandwiched operators $\hat{Q}_f=\hat{O}_T^\dagger \hat{O}_R$ and $\hat{Q}_s=\hat{O}_R^\dagger \hat{O}_R$. In the ideal case of $\hat{O}_R=\hat{O}_T$, both of these operators $\hat{Q}_s$ and $\hat{Q}_f$ reduce to the identity operator $\hat{\mathds{1}}$. We can therefore discern the quality of the operation by looking at how far we are from this ideal scenario. This analysis is best accomplished by considering the joint eigenbasis of the commuting operators $\hat{A}$ and $\hat{B}$ consisting of states $\ket{m}$ with the respective eigenvalues $m_A$ and $m_B$. Note that the basis does not need to be discrete.
We can write the diagonal elements of $\hat{Q}_f$ and $\hat{Q}_s$ as
\begin{align}\label{FPrelation}
&\bra{m}\hat{Q}_s\ket{m}=\left|\bra{m}\hat{Q}_f\ket{m}\right|^2,
   \end{align}
where the unitarity of the operator $\hat{O}_T$ is utilized. We can notice an interesting behavior: the fidelity and the success probability are not complementary and can approach unity {\em simultaneously}. In the limit of small $\tau$, the probability of success is quantified as
\begin{align}
&\bra{m}\hat{Q}_s\ket{m}\approx 1-4 m_A^2 m_B^2
   \left(m_A^2+m_B^2\right) R \tau ^6,\label{successprobability}
\end{align}
which shows the exact boundaries in the Hilbert space which supports the operation with a sufficient quality. Specifically, an approximate operation with conditional fidelity $F_c \gtrapprox 1-\epsilon$ and success probability $P_s \gtrapprox  1-\epsilon$, where $\epsilon \ll 1$, can be realized for states fully contained in Hilbert space for which $m_{max}^6 < \epsilon/(8 R \tau^6)$, where $m_{max} = \max(|m_A|,|m_B|)$. We can also rewrite the conditions in terms of the fixed total interaction strength $T = 2R\tau^2$ as:
\begin{equation}\label{m_max}
    m_{max}^6 < \frac{\epsilon R^2}{T^3},
\end{equation}
which tells us that large number of repetitions $R$ can enlarge the available support of the operation. It should also be noted that the operators $\hat{A}$ and $\hat{B}$ typically represent position, momentum, or number of quanta of the oscillators whose statistical distribution are asymptotically vanishing outside a certain range, and therefore are reasonably bounded in realistic physical systems.

The prominent aspect of our scheme is that its success probability can approach one even for many repetitions, implying that the measurement can be removed from the setup. We therefore follow the deterministic scheme depicted in Fig.~\ref{Setup}.
Formally, a single step of the operation is no longer represented by an operator $\hat{O}_1$, but by a trace preserving map which {\em deterministically} transforms any input state $\hat{\rho}_\mathrm{in}$ into
\begin{align}
&\hat{\rho}_\mathrm{out}=\mathrm{Tr}_q[\hat{U}_{xyxy}\{\ket{g}_q\bra{g}\otimes \hat{\rho}_\mathrm{in}\}\hat{U}_{xyxy}^\dagger]=\hat{O}_1 \hat{\rho}_\mathrm{in}\hat{O}_1^\dagger+\hat{O}_2 \hat{\rho}_\mathrm{in}\hat{O}_2^\dagger,\label{eq:map}
\end{align}
where  $\hat{O}_1=\bra{g}\hat{U}_{xyxy}\ket{g}=1-2\sin^2[\tau\hat{A}]\sin^2[\tau\hat{B}]+i\sin[2\tau \hat{A}]\sin[2\tau \hat{B}]/2$ is the successful operation and $\hat{O}_2=\bra{e}\hat{U}_{xyxy}\ket{g}=-\sin^2[\tau\hat{A}]\sin[2\tau\hat{B}]+i\sin[2\tau \hat{A}]\sin^2[\tau\hat{B}]$ is the erroneous operation. When the individual operation is repeated $R$ times, the final output state can be expressed as
\begin{equation}\label{}
    \hat{\rho}_\mathrm{out} = P_s \hat{O}_R \hat{\rho}_\mathrm{in} \hat{O}_R^{\dag} + (1-P_s) \hat{\rho}_\mathrm{error}^R,
\end{equation}
where $P_s$ denotes the success probability of the probabilistic scheme with otherwise identical parameters and the density matrix $\hat{\rho}_\mathrm{error}^R$ groups together all the realizations which would be in the probabilistic scenario disqualified by measurements. For states from Hilbert space limited by (\ref{m_max}) the fidelity is lower bounded by $F\ge P_s F_c \approx 1-2 \epsilon$. This result shows that the performance of the deterministic scheme is comparable to the probabilistic regime. Considering (\ref{m_max}) and the respective fidelities, the deterministic scheme achieves the performance of the probabilistic one when the number of repetitions $R$ is increased by a factor of $\sqrt{2}$.

\section*{Example of self-Kerr quantum interaction}

Let us explicitly demonstrate the performance of the proposed gate by realizing some of the nonlinear gates prevalent in quantum information theory and quantum technology. The self-Kerr operation \cite{TurchettePRL1995ConditionalPhase,LloydPRL1999} is realized by a unitary operator $\exp(iT \hat{n}^2)$ and in our approach it can be straightforwardly achieved by setting  $\hat{A}=\hat{B}=\hat{n}$, where $\hat{n}=\hat{a}^\dagger\hat{a}$ is the number operator for harmonic oscillator.  The implementation requires coupling with Hamiltonian $H \propto \hat{n}\hat{\sigma}_j$, where $\sigma_j$ are Pauli matrices. It can be obtained as part of the dispersive interaction available between two-level systems and oscillators in cavity field and membrane~\cite{ThompsonNat2008mechanical}, atoms~\cite{BocaPRL2004Dispersiveatom,GuerlinNat2007Dispersiveatom}, circuit QED~\cite{BlaisPRA2004cQEDsupcond} and superconducting systems~\cite{SchusterNat2007Superconducting}. 
In contrast to the approach of circuit QED \cite{parameterCQED}, which employs suitable time-dependent driving of the qubit-oscillator, our method employs a set of identical elementary gates, which can be repeated in order to obtain strong interaction. As a consequence, the whole operation is less demanding from the point of view of the ability to control the employed quantum systems. 
The performance of the gate can be generally estimated from the parameters and from the available dimension given by (\ref{m_max}). However, such a bound may be too loose, and actual performance depends on the specific choice of the states. Let us apply the self-Kerr operation to a sample coherent state $\ket{\beta}=\exp[\beta\hat{a}^\dagger-\beta^*\hat{a}]\ket{0}$ with $\beta = 1$.
The self-Kerr operation is non-classical and non-Gaussian operation, and produces a non-classical and non-Gaussian state when applied to a coherent state~\cite{nonclassicality}. Such states are necessary for advanced application of quantum information processing such as quantum computation \cite{Qcompcondition}, and can be recognized by negative regions of their Wigner functions \cite{KokPRA2008}. In relation to the self-Kerr effect a larger Kerr interaction strength $T$ produces more complex structures of negative Wigner function~\cite{StobinskaPRA2008WignerSelfKerr}.

\begin{figure}[ht]
\centering
\includegraphics[width=450px]{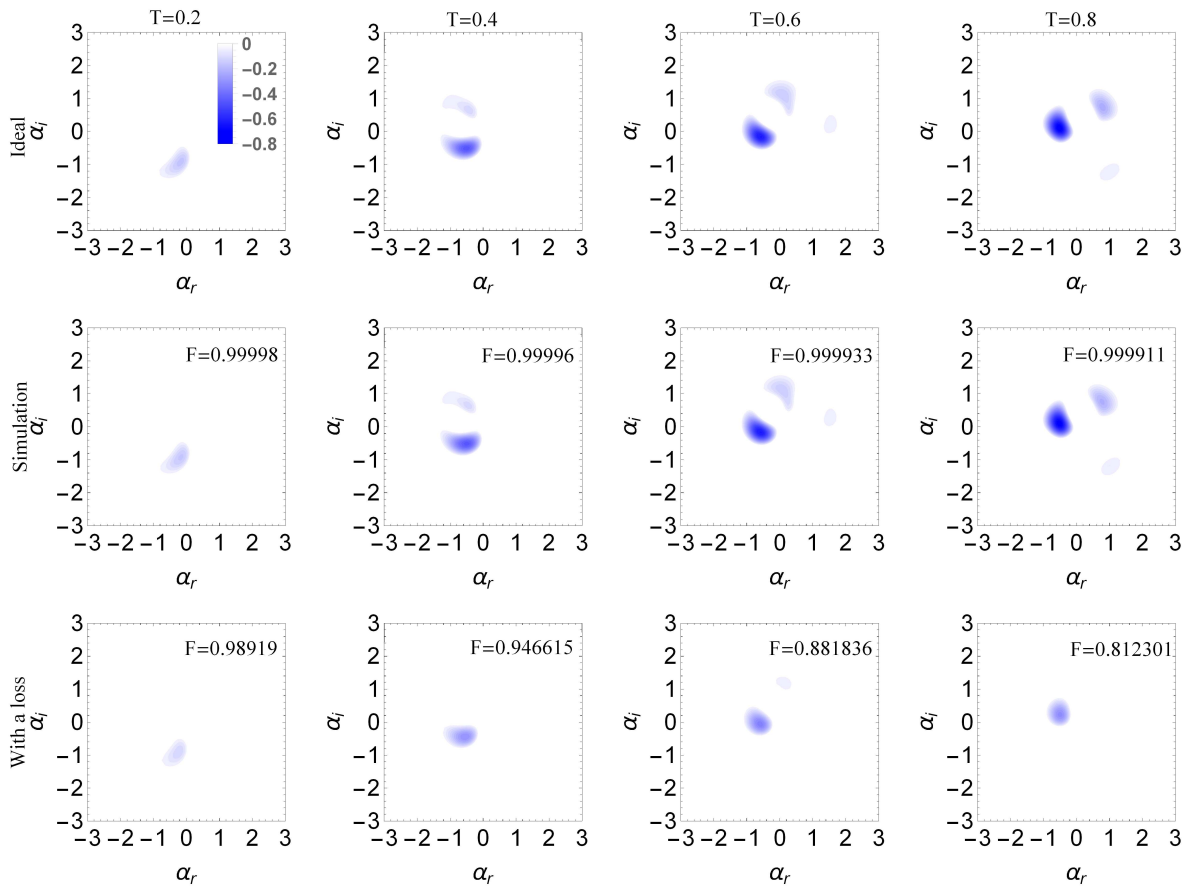}
\caption{Negative regions of Wigner functions for coherent state $\ket{\beta=1}$ subjected to self-Kerr interaction with total strengths $T=0.2$ (first column), $T=0.4$ (second column), $T=0.6$ (third column), and  $T=0.8$ (fourth column). The top row shows the ideal realization of the operation, the middle row shows simulations with single step strength of $\tau = 0.02$, and the bottom row shows realistic lossy simulation with repeated single step transmittance  $\eta=1-5.6\times 10^{-4}$. Insets show fidelities of the states with the ideal versions. We can see that the simulations faithfully recreate the ideal Wigner functions, even under the effects of moderate loss.  
}
\label{neg}
\end{figure}
In Fig.~\ref{neg}, we display the negative regions of Wigner function of self-Kerr transformed coherent states with various coupling parameters $T=0.2,0.4,0.6,0.8$. Apparently, a birth of highly nonclassical quantum interference in phase space can be observed. It is manifested by three separated regions of negativity. The figures show practically no difference between the ideal operation (above) and the deterministic approximate realization with $\tau=0.02$ (middle). This observation is reinforced by a near unit fidelity $F = 1- 0.8 \times 10^{-4}$ for $T=0.8$. Interestingly, based on (\ref{m_max}) and the parameters of the operation, the maximal Fock number allowing such high value of fidelity would be $n_{max} = 0$. As only around one third of the considered coherent state lives in that subspace, this tells us that for practical states the conditions for successful approximation might be even more relaxed. For example, for large coherent states with $|\beta|^2 \gg 1$, the fidelity of the deterministic approximative scheme scales as $F\approx 1 - 9 T^3 |\beta|^{10}/R^2$, 
derived for the lowest order expansion in $T$ of the fidelity. 
In realistic scenarios, the operation will have to endure the effects of imperfections, mainly the loss which is the dominant decoherence model for quantum oscillators. The loss can be modeled by passively coupling the evolving system to a set of zero temperature oscillators. In our model, we consider a sequence of discrete couplings, one after each cycle of the elementary sequence (\ref{eq:mix}). Each of these couplings transforms annihilation operator of the system as $\hat{a} \rightarrow \sqrt{\eta} \hat{a} + \sqrt{1-\eta}\hat{a}_{\mathrm{bath}}$, where $\hat{a}_{\mathrm{bath}}$ is annihilation operator of the auxiliary zero temperature oscillator which is immediately discarded. The single step transmittance parameter $\eta$ strongly impacts the performance of the method. To see how, we have simulated the realistic operation for $\eta=1-5.6\times 10^{-4}$. The loss counteracts the effects of the nonlinear operation. As time of the interaction increases, the state is continuously becoming more and more non-classical, which is witnessed by appearance of negative areas in its Wigner function. However, the loss is accumulated with time and at some point so much of the energy is lost that the non-classical features vanish. This can be seen in the bottom row of Fig.~\ref{neg}. We can see that while the loss of $13\%$ of the energy for $T=0.2$ did not severely affect the non-classicality, $40 \%$ loss for $T=0.8$ already removed one area of negativity. We therefore conclude that proposed method is not critically sensitive to basic decoherence caused by a loss in the oscillator.

\section*{Example of cross-Kerr quantum interaction}

Another example of quantum nonlinear interactions is the cross-Kerr coupling between two harmonic oscillators. This gate is a key component in building important two-qubit single photon gates in linear optical quantum computation such as controlled NOT gates and Fredkin gates~\cite{ChuangPRA1995Controlledgate,NemotoPRL2004CNOT, MilburnPRL1989Fredkin}, and nondestructive photon detection~\cite{ImotoPRA1985,MunroPRA2005QNDPNR}. It also enables direct photon-photon interaction used for many quantum information processing such as a one-way computation~\cite{HutchinsonJMO2004KerrOneway}. The cross-Kerr interaction, represented by a unitary operator $\exp[iT\hat{n}_1\hat{n}_2]$, can be engineered from the same fundamental component as the self-Kerr operation: the dispersive coupling between an oscillator and a qubit, only this time the qubit is coupled to two separate oscillators (as in Fig. 1b) so $\hat{A}=\hat{n}_1$ and $\hat{B}=\hat{n}_2$.

An elementary application is altering phase of a single photon based on the presence or absence of another, which is the basis for many discrete computation gates \cite{ChuangPRA1995Controlledgate,NemotoPRL2004CNOT, MilburnPRL1989Fredkin, KokRMP2007Linear}. In an example of the control-Z gate \cite{KokRMP2007Linear}, a separable state of two oscillators $\ket{00} + \ket{01} + \ket{10} + \ket{11}$ is changed to entangled state $\ket{00} + \ket{01} + \ket{10} - \ket{11}$ by the cross-Kerr gate with a strength $T=\pi$. Within our approach, the deterministic cross-Kerr gate with fidelity $F = 1-10^{-5}$ can be achieved from $R=1000$ instances of the basic block. This scenario suits the approximation well due to a limited number of photons in the systems.

\begin{figure}[ht]
\centering
\includegraphics[width=300px]{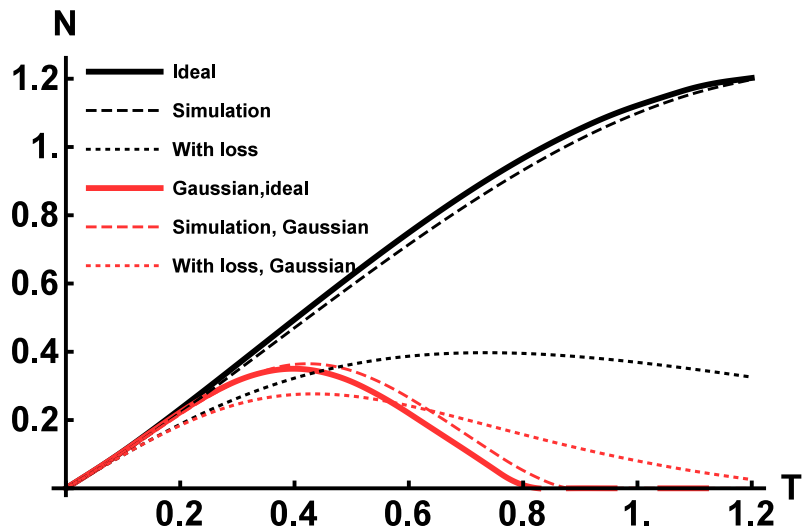}
\caption{Entanglement generated by cross-Kerr gates with different strength $T$ on a pair of coherent states $\ket{\alpha}_1\ket{\beta}_2$ with ideal cross-Kerr operator and the one achieved by our method with $\tau=0.05$. We can see that for $T>0.8$, the entanglement is purely non-Gaussian. When both oscillators suffer from loss with $\eta=1-3.5\times 10^{-3}$, we observe both reduction of overall entanglement and increase of Gaussian entanglement. This is the consequence of the loss drawing both states towards the pointer Gaussian vacuum state and Gaussifying them in the process. However, even under the effects of loss, purely non-Gaussian entanglement can still be obtained.  }
\label{CrossKerrNeg}
\end{figure}
However, there are other applications in which larger photon numbers are significant \cite{ImotoPRA1985,MunroPRA2005QNDPNR}. To test for this scenario, we consider the cross-Kerr coupling between two coherent states with amplitudes $\alpha = \beta = 1$. Considering again interaction strength $T = \pi$, the operation can be implemented with fidelity $F=0.989$ for $R=1000$ and $F = 1-5\times 10^{-4}$ with $R = 2500$ repetitions. A higher number of individual operations is demanded by the larger Hilbert space of the states for a fidelity comparable with the previous example. We can also analyze the operation from the point of view of entanglement it generates. There are several measures of entanglement \cite{entanglement}, and here we adopt the negativity due to the ease of its evaluation \cite{Vidal}. The negativity of a bipartite state given by a density operator $\rho$ can be obtained as $N[\rho]=\frac{\mathrm{Tr}[|\rho^\mathrm{PT}|]-1}{2}$ as the measure of entanglement, where $\rho^\mathrm{PT}$ is the partial transposed density matrix and $\mathrm{Tr}[|\cdot|]$ is the trace norm. The analysis should also clearly show that the cross-Kerr gate is non-Gaussian and the created entanglement should therefore be of the non-Gaussian nature. To that end we also look at the Gaussian negativity $N_G[\rho]=\frac{\mathrm{Tr}[|\rho_G^\mathrm{PT}|]-1}{2}$, where $\rho_G$ is the density matrix of a Gaussian state which has all first and second moments of quadrature operators identical with $\rho$ \cite{Laurat2005Gaussian}. Both the Gaussian and the non-Gaussian entanglement of the state generated by the cross-Kerr gate are plotted in Fig.~\ref{CrossKerrNeg} for various values of the interaction strength $T$. The interaction strength of dispersive interactions was chosen as $\tau=0.05$. We can see that the entanglement created for larger values of $T$ is practically completely non-Gaussian, as expected, and that the simulated process closely follows the ideal scenario.

To assess an impact of the decoherence on the cross Kerr interaction, we introduce an equal loss in the both oscillators.  Simulations with a realistic loss with $\eta=1-3.5\times 10^{-3}$, corresponding to the same level of noise as in a previous section, show results conceptually similar to the self-Kerr case. Again, the loss limits the achievable number of elementary gates and the corresponding total interaction strength. State with dominantly non-Gaussian entanglement can be still achieved, but the maximal difference between non-Gaussian and Gaussian entanglement is limited. For our simulation, this difference  $\max_\rho \{N[\rho]-N_G[\rho]\}$ was $0.31$ at the energy loss of about $40\%$ for a single arm.  There is, however, another interesting effect. In addition to reducing the overall correlations, the loss also drives the quantum state towards Gaussianity. As a consequence, there is less of entanglement, but higher portion of it is Gaussian. In fact, for certain values of parameters the lossy scenario produces more Gaussian entanglement than the ideal one, while non-Gaussian nature is still accessible.
It supports previous statements about a sufficient robustness of the method to the loss in oscillator.

\section*{Applications and outlook}

In summary, using a single qubit as a recyclable mediator allows for synthesis of high order nonlinear operations on quantum oscillators. These operations can be realized at an arbitrary strength with both fidelity and probability of success approaching one. The only cost is represented by the required number of repetitions of the basic building block, which may be mitigated by using an optimized architecture. Operations which can be implemented depend on the available qubit-oscillator couplings. With the feasible dispersive coupling~\cite{BlaisPRA2004cQEDsupcond,GleyzesNat2007dispersiveatom,GuerlinNat2007Dispersiveatom,SchusterNat2007Superconducting,ThompsonNat2008mechanical,JohnsonNatPhys2010QND,BocaPRL2004Dispersiveatom} it is possible to realize self-Kerr and cross-Kerr operations, which play a significant role in quantum information processing, with high quality under a moderate level of environmental effects.
The extension of the scheme ranges from engineering high order quadrature nonlinear operators, such as cubic-phase gate operator by Rabi interactions~\cite{Rabi2,Rabi3,Rabi4}, to hybrid interaction operator such as principally nonlinear optomechanical interactions~\cite{Optomechanics1,Optomechanics2,Optomechanics3} by combination of the dispersive and Rabi interactions. The higher-order versions of both dispersive and Rabi interactions open a broad class of CV nonlinear interactions.  The involved harmonic oscillators can be physically varied (optical, mechanical, electrical, collective spins), and therefore this method can potentially provide wide class of nonlinear gates between these platforms. All of these potential applications open up a possibility of deterministic quantum simulators.

\section*{ACKNOWLEDGMENT}
We acknowledge Project GB14-36681G of the Czech Science Foundation. K.P. acknowledges support by the Development Project of Faculty of Science, Palack\'y University.
\appendix
\section*{Author contributions statement}
K.P. conceived the theory. P.M. and R.F. conceived the quantification, interpreted the implications and extended the scope. P.M. and R.F. led the project.  All authors analyzed the results, wrote the article, and reviewed the manuscript.
\section*{Additional information}
\textbf{Supplementary information}  accompanies this paper at doi:;
\textbf{Competing financial interests} The authors declare that they have no competing interests.
\end{document}